\documentclass[english,10pt,aps,prd,a4paper,preprintnumbers,floatfix,nofootinbib,showpacs,superscriptaddress, twocolumn]{revtex4-1}
\usepackage[english]{babel}
\usepackage{amsmath}
\usepackage{graphicx}
\usepackage{color}
\usepackage{mathrsfs}   
\usepackage{amssymb}
\usepackage{hyperref}
\usepackage[normalem]{ulem} 
\usepackage{dcolumn}
\usepackage{bm}
\usepackage{graphicx}
\usepackage[sort&compress]{natbib}

\definecolor{greenLinks}{rgb}{0, 0.6, 0} 
\definecolor{blueLinks}{rgb}{0, 0, 0.6}
\definecolor{redLinks}{rgb}{0.6, 0, 0}
\definecolor{tempText}{rgb}{0.55, 0.10,0.67}
\definecolor{eprintLinks}{rgb}{0.4, 0.4, 0.4}
\definecolor{journalLinks}{rgb}{0.6, 0, 0}

\newcommand{\MYhref}[3][redLinks]{\href{#2}{\color{#1}{#3}}}%

\usepackage{float}

\newcommand{\AddrAHEP}{%
  AHEP Group, Institut de F\'{i}sica Corpuscular --
  C.S.I.C./Universitat de Val\`{e}ncia, Parc Cient\'ific de Paterna.\\
 C/ Catedr\'atico Jos\'e Beltr\'an, 2 E-46980 Paterna (Valencia) - SPAIN}

\begin{document}

\title{Can one ever prove that  neutrinos are Dirac particles? }
\author{Martin Hirsch}\email{mahirsch@ific.uv.es}
\affiliation{\AddrAHEP}
\author{Rahul Srivastava}\email{rahulsri@ific.uv.es}
\affiliation{\AddrAHEP}
\author{Jos\'{e} W. F. Valle}\email{valle@ific.uv.es}
\affiliation{\AddrAHEP}

\begin{abstract}
   \vspace{1cm}
   
   According to the ``Black Box'' theorem the experimental
   confirmation of neutrinoless double beta decay ($0 \nu 2 \beta$)
   would imply that at least one of the neutrinos is a Majorana
   particle. However, a null $0 \nu 2 \beta$ signal cannot decide the
   nature of neutrinos, as it can be suppressed even for Majorana
   neutrinos. In this letter we argue that if the null $0 \nu 2 \beta$
   decay signal is accompanied by a $0 \nu 4 \beta$ quadruple beta
   decay signal, then at least one neutrino should be a Dirac
   particle. This argument holds irrespective of the underlying
   processes leading to such decays.
   
\end{abstract}


\maketitle

Ever since the early days of neutrino
physics~\cite{Fermi:1934hr,GoeppertMayer:1935qp,Majorana:1937vz,Furry:1939qr}
there has been a debate about the nature of neutrinos i.e. whether
they are Dirac or Majorana fermions.
The debate has origins in the fact that, although most of the known
fermions (except neutrinos, whose nature is yet to be ascertained) are
Dirac particles and hence four-component spinors, the fundamental
irreducible spinorial representations of the Poincar\'{e} group are
actually two-component. 
However, the Poincar\'{e} group describes just the kinematics, and
does not represent the full unbroken symmetry of nature.

Apart from spacetime symmetry, particle theories also have ``internal
symmetries'' for example ``gauge symmetries'', such as the
$SU(3)_C \otimes SU(2)_L \otimes U(1)$ of our cherished Standard Model
(SM).
According to the gauge paradigm, these symmetries dictate the dynamics
of all fundamental processes amongst elementary particles.
The SM gauge group is spontaneously broken by the celebrated
Brout-Englert-Higgs mechanism, but not completely. As far as we know
from experiments, an $SU(3)_C \otimes U(1)_{EM}$ gauge symmetry
remains unbroken.
This symmetry then dictates the dynamics of fundamental processes at
energies below the electroweak symmetry breaking scale.
Thus at energy or temperature scales well below the electroweak
breaking scale, one must not only take into account the invariance
under the Poincar\'{e} group, but also under the unbroken
$SU(3)_C \otimes U(1)_{EM}$ gauge group.
Thus, any fermion carrying a non-zero color or electric charge cannot
have a Majorana mass term, since such term would necessarily break the
$SU(3)_C \otimes U(1)_{EM}$ gauge symmetry.
This implies that, although two-component spinors are indeed
fundamental, the requirement that color and electromagnetic charges
remain conserved, forces all the quarks and charged leptons to be Dirac
particles.
On the basis of this argument it has been argued
in~\cite{Schechter:1980gr} that, thanks to their complete charge
neutrality, only neutrinos can be -- and should be -- Majorana
fermions.
However, nature need not follow our theoretical prejudices, so that
only experiments can settle whether neutrinos are Dirac or Majorana
particles.

Thanks to the small neutrino mass $m_\nu$ and the V-A nature of the
weak interaction, discerning the nature of neutrinos from experiments
is a formidable task. A basic difference between Dirac and Majorana
fermions resides in the CP phases present in their mixing
matrices~\cite{Schechter:1980gr}. Indeed the sensitivity to the
physical Majorana phases present in neutrino to anti-neutrino
oscillations~\cite{Schechter:1981gk} is well below any conceivable
test. Likewise, electromagnetic properties of
neutrinos~\cite{Schechter:1981hw,Li:1981um,kayser:1982br} have a
hidden dependence on $m_\nu$. Indeed, all observables sensitive to the
Majorana nature of neutrinos end up being suppressed by a power of
$m_\nu$.
The small scale of the active neutrino masses makes such differences
very tiny.

\begin{figure}[!h]
 \centering
  \includegraphics[scale=0.25]{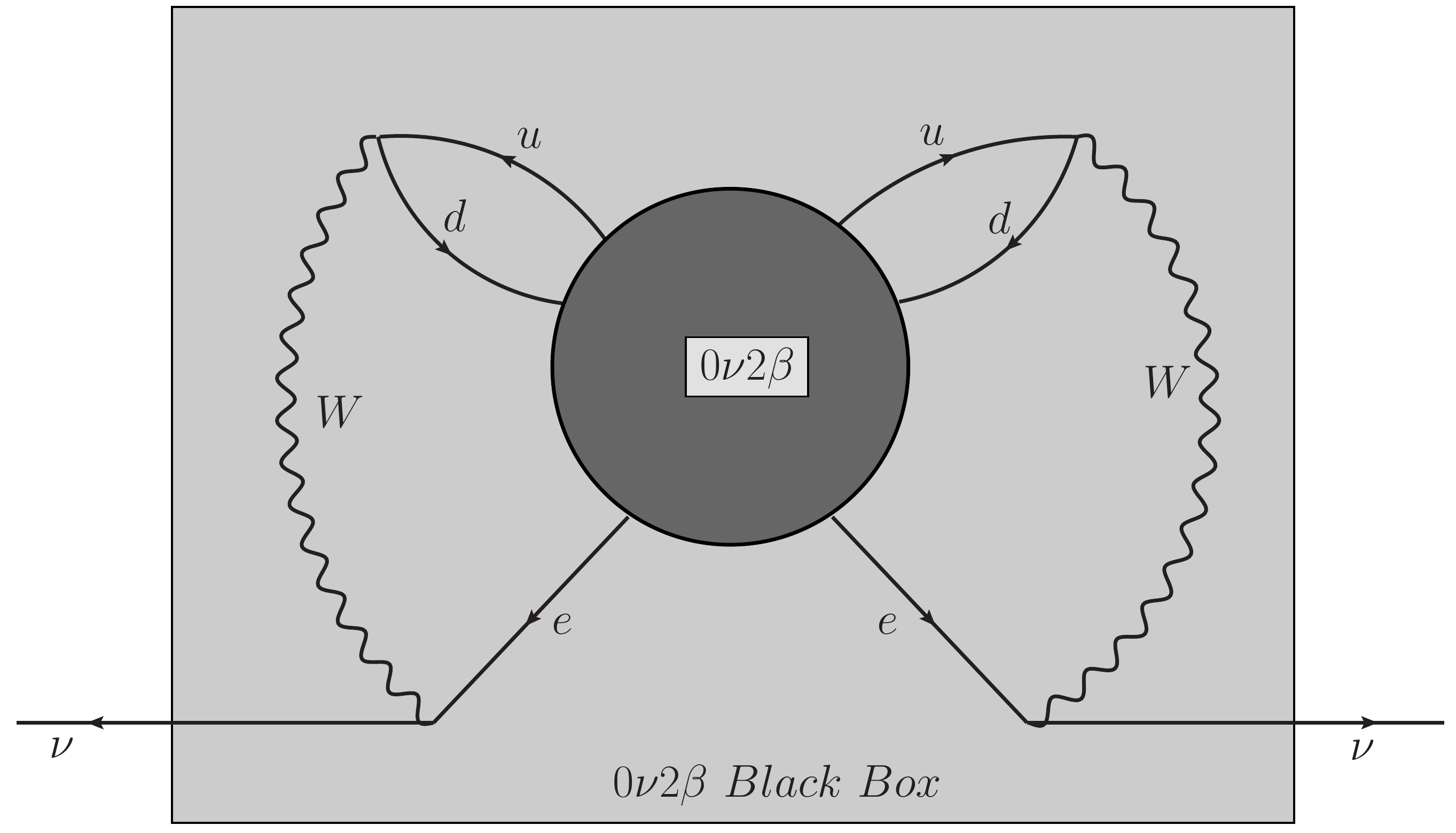}
  \caption{The ``Black Box'' theorem states that a $0\nu2\beta$ signal
    ensures that at least one neutrino is Majorana in
    nature~\cite{Schechter:1981bd}.  }
    \label{black-box}
  \end{figure}

However, there is a potentially feasible process which may settle the
issue, namely the neutrinoless double beta decay, which has long been
hailed as the ultimate test concerning the nature of
neutrinos~\footnote{In SM extensions there may be feasible
  complementary probes of lepton number violation at collider
  energies~\cite{Rizzo:1982kn,keung:1983uu,Deppisch:2012nb,Das:2012ii,Deppisch:2013jxa}.}.
Indeed, if $0 \nu 2 \beta$ decay is ever observed, its amplitude can
always be ``dressed'' so as to induce a Majorana mass, ensuring that
at least one of the neutrinos is of Majorana
type~\cite{Schechter:1981bd}, as illustrated in Fig.~\ref{black-box}.
See Ref.~\cite{hirsch:2006yk,Duerr:2011zd} for recent discussions.

  However, the non-observation of $0 \nu 2 \beta$ decay so far
  \cite{Agostini:2017iyd,Albert:2014awa,KamLAND-Zen:2016pfg,Alduino:2017ehq}
  has raised the intriguing possibility that neutrinos might well be Dirac
  particles.
  Several well motivated high-energy completions of the SM do lead to
  naturally light Dirac-type
  neutrinos~\cite{Chen:2015jta,Addazi:2016xuh,Valle:2016kyz,Reig:2016ewy}.
  Alternatively, the absence of a $0 \nu 2 \beta$ signal is not
  inconsistent with the Majorana nature of neutrinos, since the decay
  amplitude may be suppressed as a result of a destructive
  interference amongst the three active neutrinos, even if they are
  Majorana type~\cite{wolfenstein:1981rk,valle:1982yw}.
  Thus, although the observation of $0 \nu 2 \beta$ decay would
  necessarily imply that at least one neutrino species is Majorana in
  nature, the converse is not true: a negative $0 \nu 2 \beta$ decay
  signal does not tell us anything about the nature of neutrinos.

  This prompts us to search for processes beyond the simplest
  $0 \nu 2 \beta$ decay which can also shed light upon the nature of
  neutrinos.\footnote{Observation of a non-zero mass in KATRIN, together
    with non-observation of $0\nu 2\beta$ decay would also favour Dirac
    neutrinos.}
  We will specifically focus on the two lowest $0 \nu 2n \beta$
  processes characterized by $n = 1,2$, namely, the neutrinoless
  double beta decay $0 \nu 2 \beta$ and the neutrinoless quadruple
  beta decay.\\[-.2cm]

  An experimental search for the $0 \nu 4 \beta$ process has been
  recently performed by the NEMO-3 collaboration, using
  $^{150}$~Nd~\cite{Arnold:2017bnh}. The possible existence of
  $0 \nu 4 \beta$ decays has been first suggested
  in~\cite{Heeck:2013rpa}, and it is expected to arise in a number of
  models with family symmetries leading to Dirac
  neutrinos~\cite{Chulia:2016giq,Chulia:2016ngi,CentellesChulia:2017koy}.
  Here we argue that the combination of the $0 \nu 2 \beta$ and
  $0 \nu 4 \beta$ processes may be enough to settle the nature of
  neutrinos within a very broad class of models.\\[-.2cm]

  In order to proceed let us first look at the $0 \nu 2 \beta$ process
  and the neutrino mass generation from the symmetry point of view.
  In the Standard Model the neutrinos are massless and there is an
  accidental global ``classically conserved'' $U(1)_L$ symmetry in the
  lepton sector associated to Lepton number for all the leptons in
  SM~\footnote{There is an additional accidental global $U(1)_B$
    symmetry associated to conserved Baryon number.  While B and L are
    separately anomalous at the quantum level, there are anomaly free
    combinations, such as $U(1)_{B-L}$.  For simplicity here we
    discuss only $U(1)_L$, though our argument remains valid for
    $U(1)_{B-L}$. }.
  By just adding right handed neutrinos $\nu_{iR}$ sequentially to the
  SM particle content one can give mass to neutrinos without breaking
  the lepton number symmetry.
  In such a case neutrinos will necessarily be Dirac particles and the
  $0 \nu 2n \beta$; $n \geq 1$ decays will all be absent.

  We now turn to the cases when this lepton number is broken down to a
  discrete $Z_m$ subgroup ($m \geq 2$) which remains conserved.
  Notice that a $U(1)$ symmetry only admits $Z_m$ subgroups, where
  $Z_m$ is a cyclic group of $m$ elements, characterized by the
  property that if $x$ is a non-identity group element, then
  $x^{m+1} \equiv x$. The $Z_m$ groups only admit one-dimensional
  irreducible representations, conveniently represented by using the
  n-th roots of unity, $\omega = e^{\frac{ 2 \pi I}{m}}$, where
  $\omega^m = 1$.
  If lepton number is broken to a $Z_m$ subgroup (with neutrinos
  transforming non-trivially under $Z_m$) by the new physics
  responsible for neutrino mass generation, then we have two possible
  cases:
\begin{eqnarray}
 U(1)_L   & \, \to  \, &   Z_m \equiv Z_{2n+1} \, \text{where} \,  n \geq 1 \, \text{ is a positive integer}    \nonumber \\
 & \, \Rightarrow \, & \text{Neutrinos are Dirac particles}      \nonumber \\
 U(1)_L  & \, \to  \,  & Z_m \equiv Z_{2n} \, \text{where} \,  n \geq 1 \, \text{ is a positive integer}  \nonumber \\
 & \, \Rightarrow \, & \text{Neutrinos can be Dirac or Majorana } 
\label{oddzn}
\end{eqnarray}
If the $U(1)_L$ is broken to a $Z_{2n}$ subgroup, then one can
make a further broad classification
\begin{eqnarray}
 \nu & \sim & \omega^{n} \,\, \text{under} \, Z_{2n} \, \Rightarrow \, \text{Majorana neutrinos} 
 \label{evenznmaj} \\
 \nu & \nsim & \omega^{n} \,\, \text{under} \, Z_{2n} \, \Rightarrow \, \text{Dirac neutrinos} 
 \label{evenzndir}
\end{eqnarray}
depending on the charges of neutrinos under the unbroken $Z_{2n}$
symmetry. For neutrinos transforming non-trivially under any
unbroken $Z_{2n + 1}$ symmetry, they must be Dirac particles. 
For neutrinos transforming non-trivially under the $Z_{2n}$
symmetry, they can be Majorana if and only if
$\nu \, \sim \, \omega^{n}$. For any other transformation neutrinos
will be Dirac particles.
Thus, from a symmetry point of view, in contrast to popular belief,
the Majorana neutrinos are the special ones, emerging only for certain
transformation properties under the unbroken residual $Z_{2n}$
symmetry.

\begin{figure}[!h]
 \centering
  \includegraphics[scale=0.25]{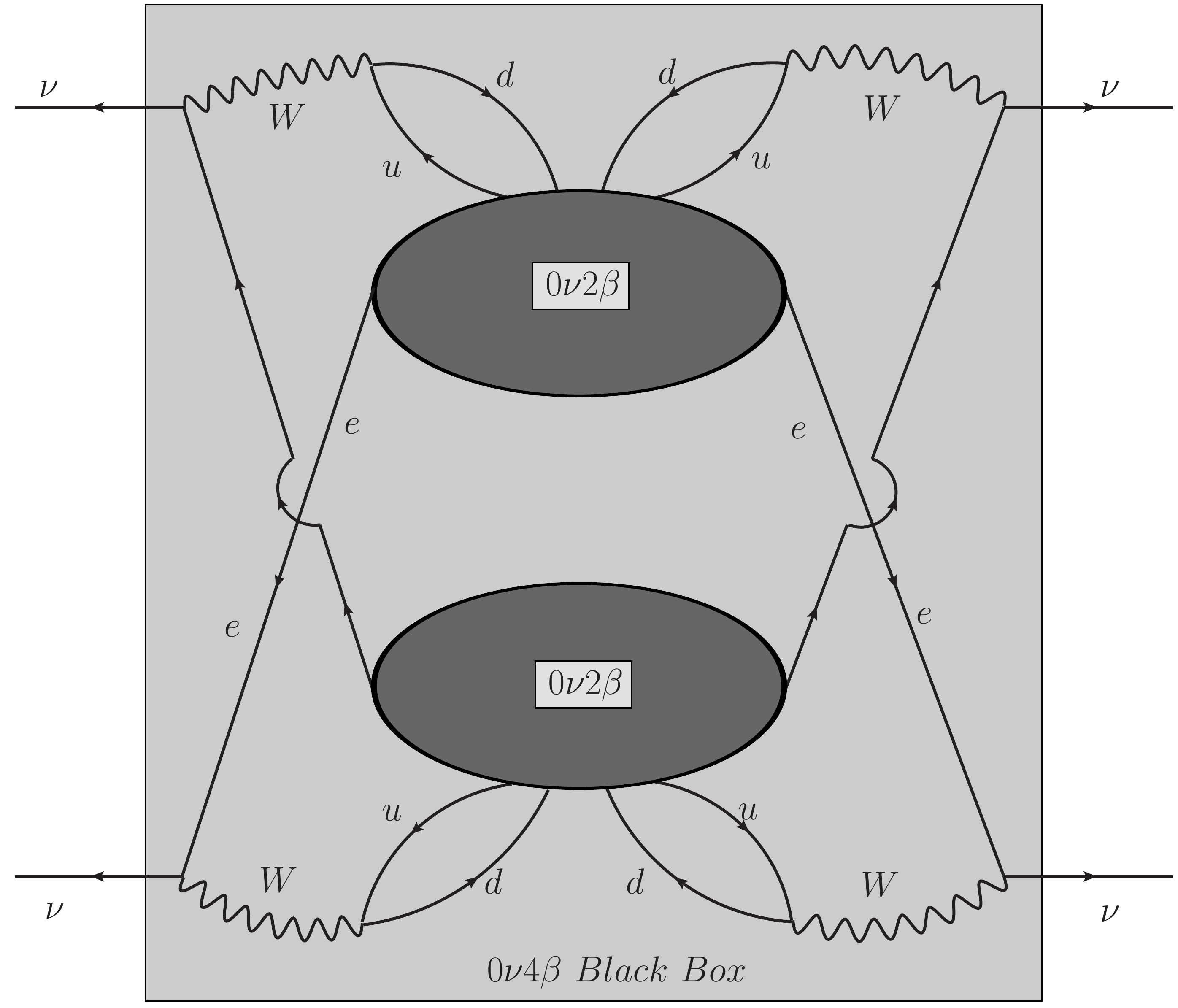}
  \caption{$0 \nu 4 \beta$ arising as a double $0 \nu 2 \beta$
    process}
    \label{fig2}
  \end{figure}

Now the simplest $Z_m$ group to which the $U(1)_L$ can break is
$Z_2$. This case is special, as it only offers two possibilities for
neutrino transformation i.e. $\nu \sim +1 \, \text{or} \, -1$, both of
which satisfy Eq.~(\ref{evenznmaj}) and only allows for Majorana
neutrinos.
Breaking $U(1)_L$ to $Z_2$ is quite simple, through a Majorana
mass term $\nu\nu$ arising effectively from new physics, as is the
case of Weinberg's dimension 5 operator
$\bar{L}^c \Phi \Phi L$~\cite{weinberg:1979sa}.
Most popular in the literature, this case covers a big chunk of model
setups, which typically involve breaking of lepton number to a
residual $Z_2$ symmetry.
This also induces a nonzero $0 \nu 2 \beta$ decay amplitude, as this
decay is now allowed by the symmetry. The converse is also true,
namely, if the $0 \nu 2 \beta$ decay process is allowed, it always
implies that lepton number is broken and the associated new physics is
bound to generate Majorana mass terms~\footnote{Notice that the
  Majorana mass term might be generated at the loop level, and need
  not be the dominant source of $0 \nu 2 \beta$ decay.}. Notice that,
since the higher $0 \nu 2n \beta$ beta process are also allowed by the
residual $Z_2$ symmetry, they all will also occur through
`multiples of n'' $0 \nu 2 \beta$ amplitudes as illustrated in
Fig.\ref{fig2}, for the simplest case of n=2.
These higher processes can be intuitively thought of as ``multiples''
of the basic $0 \nu 2 \beta$ process,
$0 \nu 2n \beta \equiv n(0 \nu 2 \beta)$ and thus we have
$\Gamma_{0 \nu 2n \beta} \ll \Gamma_{0 \nu 2 \beta}$.

We now turn to the case of $U(1)_L$ broken to higher symmetries, with
neutrinos transforming non-trivially under the residual $Z_m$
symmetry.
Clearly if $U(1)_L$ breaks to an $Z_{2n + 1}$ symmetry, the lowest
possible allowed neutrinoless beta decay process will be $0\nu (2n +
1) \beta$, where $n$ is a $\text{positive integer}$. But such
processes are forbidden, as can be easily seen. Consider, for
simplicity $0\nu 3\beta$. This process would require us to write down
a 9-fermion operator, which is of course not possible.  \footnote{
  Notice that although the lowest $0\nu (2n + 1) \beta$ is forbidden,
  the higher dimensional $0\nu a(2n + 1) \beta$ processes ($a$ is a
  even integer and $a \geq 2$) are still allowed by $Z_{2n + 1}$
  symmetry.}. Hence in such cases no neutrinoless beta decay of any
order below $0\nu 2(2n + 1) \beta$ are possible and neutrinos can only
be Dirac particles~\cite{Ma:2015mjd,Bonilla:2016diq}.

  The more interesting case is when $U(1)_L$ breaks to even residual
  $Z_{2n}$ symmetries, with $n>1$. As already mentioned, in such cases
  both Dirac and Majorana neutrinos are possible, depending on how
  they transform under the $Z_{2n}$ symmetry. 
  Also, irrespective of the Dirac or Majorana nature of neutrinos, if
  $U(1)_L$ breaks to an even residual $Z_{2n}$ symmetry, there is an
  associated $0\nu 2n \beta$ processes allowed by the residual
  symmetry.
  However, an important distinction comes for the case of Dirac or
  Majorana neutrinos. As mentioned above, if neutrinos transform as
  $\omega^{n}$ under the $Z_{2n}$ symmetry, they must be Majorana
  particles. 
  Moreover, in this case not only the $0\nu 2n \beta$ process is
  allowed, but all other lower dimensional $0\nu 2n_1 \beta$
  processes, where $n_1 < n$ is a positive integer, are also allowed.
  However, if neutrinos are Dirac particles, then for the case of a
  $Z_{2n}$ symmetry, it follows that $\nu \nsim \omega^{n}$. This
  implies that the lowest process allowed by $Z_{2n}$ symmetry is
  $0 \nu 2n \beta$ decays, all other lower dimensional processes being
  forbidden by the unbroken residual $Z_{2n}$ symmetry.

\begin{figure}[!h]
 \centering
  \includegraphics[scale=0.25]{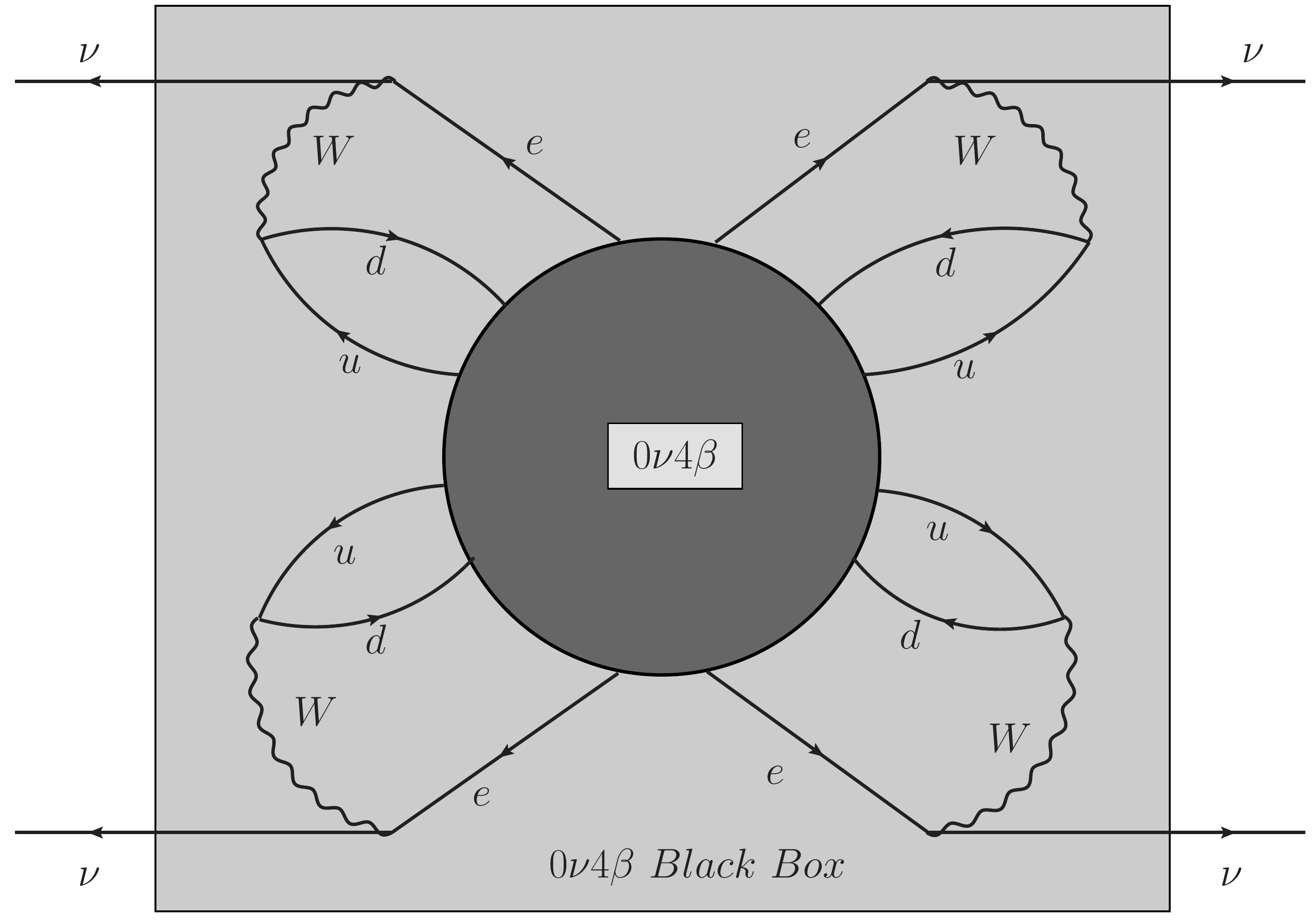}
  \caption{The quadruple beta decay process is allowed by a residual
    $Z_4$ symmetry irrespective of the nature of neutrinos.}
    \label{quad-beta}
  \end{figure}
  This is better illustrated by the simple example of $U(1)_L$
  breaking to a $Z_4$ residual symmetry, called quarticity. Such a
  breaking has been accomplished within concrete realistic gauge
  models~\cite{Heeck:2013rpa,Chulia:2016giq,Chulia:2016ngi,CentellesChulia:2017koy}. As
  already mentioned, in this case both Dirac as well as Majorana
  neutrinos are possible. Neutrinos will be Dirac if they transform as
  $\omega$ or $\omega^3$; with $\omega^4=1$. They will be Majorana
  otherwise, if they transform trivially or transform as $\omega^2$
  under $Z_4$.  However, the quadruple beta decay $0 \nu 4 \beta$
  illustrated in Fig. \ref{quad-beta} is always allowed, irrespective
  of the nature of neutrinos.

  Notice that, if neutrinos are Majorana particles transforming as
  $1, \omega^2$ under the $Z_4$ symmetry, the lower dimensional
  $0 \nu 2 \beta$ diagram of Fig. \ref{black-box} is also allowed by
  the $Z_4$ symmetry.  
  By dimensional power counting one sees that the $0 \nu 2 \beta$
  decay is induced by a dim-9 operator, whereas $0 \nu 4 \beta$ decay
  is induced by dim-18 operator.
  Barring extremely fine tuned cancellations, one naively expects
  that $\Gamma_{0 \nu 2 \beta} \gg \Gamma_{0 \nu 4 \beta}$.
  We estimate $R=\Gamma_{0 \nu 2 \beta}/\Gamma_{0 \nu 4 \beta}$
    for two simple cases: (a) $0 \nu 2 \beta$ and $0 \nu 4 \beta$
    induced by effective $d=9$ and $d=18$ operators, respectively.
    And, (b) $0 \nu 2 \beta$ induced by a Majorana neutrino mass,
    while $0 \nu 4 \beta$ is induced by the ``lepton quarticity'' $d=6$
    operator. We find for case (a):
    \begin{equation}\label{eq:RatOp}
      R = \frac{Q_{\beta\beta}^5(\frac{1}{\Lambda^5})^{2}q^6}
           {Q_{4\beta}^{11}(\frac{1}{\Lambda^{14}})^{2}q^{18}}
           \sim 10^{82},
    \end{equation}       
    while case (b) gives:
     \begin{equation}\label{eq:RatMM}
      R = \frac{Q_{\beta\beta}^5(\frac{G_F^2(Y_{\nu}v)^2}{\Lambda})^{2}q^2}
          {Q_{4\beta}^{11}(\frac{G_F^4}{\Lambda^{2}})^{2}q^{10}}
           \sim 10^{30}.
    \end{equation}       
 Here $Q_{\beta\beta}$ and $Q_{4\beta}$ are the Q-values of
      the decays, both of order MeV.  $G_F$ is the Fermi constant,
      $q \simeq 0.1$ GeV is the typical momentum transfer in the
      nucleus and $\Lambda$ is the scale characterizing the new
      physics. The numbers correspond to $\Lambda \sim 1$ TeV and for
      the mass mechanism we have included that the resulting neutrino
      mass should be at the level of the current experimental bound. 
      
  Thus for Majorana neutrinos one naively expects to
  first see neutrinoless double beta decay, if at all.
  In contrast, for Dirac neutrinos, the dim-18 neutrinoless quadruple
  beta decay process is still allowed by $Z_4$ symmetry, while
  ``conventional'' neutrinoless double beta decay process is
  forbidden.
  Therefore, barring exceptional cases, if future experiments were to
  observe neutrinoless quadruple beta decay~\cite{Arnold:2017bnh}
  without a positive neutrinoless double beta decay signal, then
  neutrinos should be Dirac particles. This conclusion can be easily
  generalized to higher $Z_{m}$ symmetries and higher $0\nu2n\beta$
  decays.

  Another important conclusion is that, for neutrinos to be Majorana
  particles, lepton number $U(1)_L$ must be broken to an even $Z_{2n}$
  subgroup under which neutrinos must transform in a very special
  way. Such possible ``special nature'' of Majorana neutrinos
  following from symmetry considerations is at odds with popular
  prejudices.
  In such a case all $0 \nu 2n \beta$ neutrinoless beta decay
  processes can be potentially induced, as they are all allowed by the
  $Z_{2n}$ symmetry. If neutrinos do not transform appropriately then
  they must be Dirac particles. 
  In such a case the lowest possible neutrinoless beta decay process
  allowed by symmetry is $0 \nu 2n \beta$ instead of the
  ``conventional'' $0 \nu 2 \beta$ decay. All lower dimensional
  neutrinoless beta decay process are forbidden by $Z_{2n}$ symmetry.
  If, by contrast, $U(1)_L$ is broken to an odd $Z_{2n + 1}$ symmetry
  with neutrinos transforming non-trivially under it, then neutrinos
  are necessarily Dirac particles and all neutrinoless beta decay
  process of any dimension lower than $0 \nu 2(2n+1) \beta$ decay 
are forbidden.
  It may also happen that $U(1)_L$ is either completely broken with no
  residual subgroup or is broken in such a way that the neutrinos
  transform trivially under the residual discrete lepton number $Z_m$
  symmetry.  In either of these cases, neutrinos can again be Majorana
  and all $0 \nu 2n \beta$ processes, including the $0 \nu 2 \beta$
  decay are allowed by symmetry. In such a scenario also, on
  dimensional grounds, one should expect to first observe
  $0 \nu 2 \beta$ decay before observing any higher $0 \nu 2n \beta$
  process.

  The whole discussion above leads to an important conclusion
  concerning the nature of neutrinos, namely, if we first observe a
  higher $0 \nu 2n \beta $ decay, unaccompanied by a lower dimensional
  neutrinoless beta decay signal, such as $0 \nu 2 \beta$ decay, that
  would be a strong indication in favour of the Dirac nature of
  at least one neutrino. 
  This statement holds in general, with the possible exception of very
  special cancellations as present say, in the quasi-Dirac neutrino
  situation~\cite{wolfenstein:1981rk,valle:1982yw}, which would
    require fine tuning at the $\sim 10^{-30}$ level. 
  In contrast, if neutrinos are Majorana particles, then we should
  first observe $0 \nu 2 \beta$ decay, if at all, well before
  observing any higher dimensional $0 \nu 2n \beta $ decays.
  In short, we have argued that, should a $0 \nu 4 \beta$ decay signal
  ever be established, unaccompanied by $0 \nu 2 \beta$ decays, then
  one would rule out Majorana neutrinos.

\begin{acknowledgments}

  We wish to thank S.C. Chuli\'{a} for helpful discussions.  Work
  supported by the Spanish grants FPA2017-85216-P and SEV-2014-0398
  (MINECO), and PROMETEOII/2014/084 (Generalitat Valenciana). The
  Feynman diagrams are drawn using Jaxodraw \cite{Binosi:2003yf}.
 
\end{acknowledgments}


\bibliographystyle{bib_style_T1}

\end{document}